\documentstyle[a4wide,11pt]{article}

\input amssym.def
\input amssym.tex
\newcommand\RR{{\Bbb R}}

\newcommand\NN{{\Bbb N}}
\newcommand\ZZ{{\Bbb Z}}

\newcommand\Dra[1]{{\cal H}_{#1}}
\newcommand\ud[1]{(#1,\Dra{#1})}
\newcommand\upi{\ud{\pi}}

\def\d={\,:=\,}

\newcommand{\semdir}
{\rtimes}

\newcommand{\rhk}
 {{\RR{}^k}}
\newcommand{\lpraum}[2]
 {{{\rm L}}^{#1}(#2)}

\newcommand{\lzw}
 {\lpraum{2}\rhk{}}
\newcommand{\rhkdu}
 {{\widehat{\rhk{}}}}

\font\frakten=eufm10
\newfam\frakfam
\textfont\frakfam=\frakten

\newtheorem{thm}{Theorem}[section]
\newtheorem{lemma}[thm]{Lemma}
\newtheorem{cor}[thm]{Corollary}
\newtheorem{prop}[thm]{Proposition}
\newtheorem{Defn}[thm]{Definition}
\newtheorem{Ex}[thm]{Example}
\newtheorem{Rem}[thm]{Remark}
\newtheorem{Exs}[thm]{Examples}
\newtheorem{Rems}[thm]{Remarks}
\newtheorem{Defrem}[thm]{Definition and Remark}
\newtheorem{Remnt}[thm]{}
\newenvironment{defn}
 {\begin{Defn} \begin{rm}} {\end{rm} \hfill $\Box$ \end{Defn}}
\newenvironment{defrem}
 {\begin{Defrem} \begin{rm}} {\end{rm} \hfill $\Box$ \end{Defrem}}

\newenvironment{rem}
 {\begin{Rem} \begin{rm}} {\end{rm} \hfill $\Box$ \end{Rem}}

\newenvironment{remnt}[1]{\begin{Remnt} {\bf #1} \begin{rm}} {\end{rm}
 \hfill $\Box$ \end{Remnt}}
\newenvironment{prf} {{\bf Proof.}}{\hfill $\Box$}

\begin{document}

\title{Continuous wavelet transforms from semidirect products: Cyclic
 representations and Plancherel measure}
\author{Hartmut F\"uhr \thanks{email: fuehr@mathematik.tu-muenchen.de},
 Matthias Mayer \thanks{email: mayerm@mathematik.tu-muenchen.de}
 \\ Zentrum Mathematik der 
 Technischen Universit\"at M\"unchen, \\D--80290 M\"unchen}
\date{\today}
\maketitle
\begin{abstract}
 Continuous wavelet transforms arising from the quasiregular representation
 of a semidirect product group $G = \RR^k \semdir H$ have been studied 
 by various authors. Recently the attention has shifted from the irreducible
 case to include more general dilation groups $H$, for instance 
 cyclic (more generally: discrete) or one-parameter groups. These
 groups do not give rise to irreducible square-integrable representations,
 yet it is possible (and quite simple) to give admissibility conditions for
 a large class of them. 
 We put these results in a theoretical context by establishing a
 connection to the Plancherel theory of the semidirect products, and show how
 the admissibility conditions relate to abstract admissibility conditions
 which use Plancherel theory.
\end{abstract}

\section*{Introduction}

In one of the initial papers of wavelet analysis \cite{GrMoPa},
Grossmann, Morlet and Paul proved
that the continuous wavelet transform on ${\rm L}^2(\RR)$ and its
inversion formula rests on
a certain representation of the semidirect product $\RR \semdir \RR^+$
acting on ${\rm L}^2(\RR)$. More precisely, they showed that the
square-integrability (in the sense of \cite{DuMo})
of that representation guarantees the existence of an inverse
wavelet transform, and they showed how the admissibility
conditions related to the so-called {\em Duflo-Moore operators},
which are naturally associated to square-integrable representations.
This realization opened the way to analogous constructions in
a variety of settings. One class of groups and representations 
attracting particular attention are the semidirect products of the 
type $\RR^k \semdir H$.
Here $H$ is a closed matrix group (the so-called {\em dilation group}).
$G$ has a natural unitary representation on ${\rm L}^2(\RR^k)$,
which is the chief object of study for the construction of
wavelet transforms. Concrete higher-dimensional examples of
such transforms were produced
by Murenzi \cite{Mu} and Bohnke \cite{Bo}. Later the problem
was studied in greater generality by Bernier and Taylor
\cite{BeTa}, who
gave sufficient conditions for the existence of inversion
formulas, and also calculated the Duflo-Moore operators for
these cases. The general approach was further pursued in 
\cite{Fu,AnCaDeLe}.

All the references cited so far restrict attention to the
case of {\em irreducible} square-integrable representations
(also called {\em discrete series representations}). But
several authors produced evidence that inversion formulas
could also be obtained for non-irreducible representations
\cite{IsKl,KlSt,MaZh}. All results were given for the
semidirect product setting, and they turned out to be 
rather simple generalizations of what had been obtained in the 
irreducible case. A more general family of dilation groups was studied by
Gr\"ochenig, Kaniuth and Taylor \cite{GKT}, who focussed on
certain one-parameter groups. Their so-called 
``projection generating functions'' turn out to be particular
admissible vectors. 
Quite recently, G. Weiss and collaborators announced in \cite{WW}
and proved in \cite{LWWW} an almost characterization of those
dilation groups which admit an inversion formula; in particular
all of the aforementioned examples fall under the class
described in \cite{LWWW}.

One intriguing aspect of the concrete results in \cite{IsKl,KlSt,MaZh}
(and the initial motivation for this paper) was that, until recently,
no general representation-theoretic framework,
comparable to the characterization for irreducible representations,
existed in which the concrete admissibility conditions would fit.
The theoretical framework has now been provided in \cite{Fu3},
where general admissibility conditions are formulated by use of
the Plancherel decomposition of the regular representation of the
underlying group. It is the main purpose of this paper to exhibit
the relationship between the concrete results on semidirect
products and the abstract
admissibility conditions, and thus to provide a bridge between
\cite{LWWW} and \cite{Fu3}.

Our paper has three sections. In the first section, we consider
semidirect products $\RR^k \semdir H$ and their quasi-regular
representations on ${\rm L}^2(\RR^k)$. We derive admissibility
criteria for subrepresentations of quasi-regular representations,
and give sufficient conditions for the existence of admissible
vectors. The results in that section have to a large extent 
been proved in \cite{LWWW}. We include proofs for several reasons:
Our results are slightly more general in that we consider arbitrary
subrepresentations of the quasiregular representations.
Secondly, the bulk of our results was obtained independently
from \cite{WW,LWWW}.
The third and most important reason is that most of the
constructions and objects used in the proof of the concrete 
results turn up again in the abstract setting; most notably
the dual orbit space $\rhkdu / H$ and certain measures
on the orbits and the orbit space. Hence including
the proofs should facilitate understanding the relationship between
abstract and concrete results.

Section two is devoted to a short review of Plancherel theory
and its relation to admissibility conditions. Section three
then connects the results of the previous two sections, by
identifying the Duflo-Moore operators and the Plancherel
measure with certain operators and measures obtained in the
concrete setting. We wish
to point out, though, that in the course of our argument
the concrete admissibility conditions do not appear as mere corollaries
of the abstract ones. We use the concrete results to show that
they are special cases of the abstract results. Hence the results
of the third section might be considered redundant, but we maintain
that it is useful to work out this class of examples
and to explicitly compute the various objects of Plancherel
theory, i.e., the Duflo-Moore operators and the corresponding
Plancherel measure.

Apart from serving as an illustration for the results in \cite{Fu3},
the discussion in Section 3 also ties in nicely with
the results of Kleppner and Lipsman \cite{KlLi} on the Plancherel
theory of semidirect products in general (see the discussion in
Section 3 for a more detailed account of their results). 
We wish to stress that while the results of Kleppner and Lipsman 
provide an orientation for our arguments in Section 3, our proof
does not use their calculation of Plancherel measure.

In the remainder of the introduction we wish to collect some generalities
concerning wavelet transforms and admissible vectors. Assume
that $\pi$ is a (strongly continuous, unitary) representation of
a locally compact group $G$ on the Hilbert space ${\cal H}_{\pi}$.
For two vectors $\eta, \phi \in {\cal H}_{\pi}$ and $x \in G$ define
\[ V_{\eta} \phi (x) := \langle \phi, \pi(x) \eta \rangle ~~.\] 
Then $V_{\eta}$ is a bounded operator from ${\cal H}_{\pi}$
into $C_b(G)$. We call $\eta$ {\bf admissible} if $V_\eta$ is
an isometry from ${\cal H}_{\pi}$ into ${\rm L}^2(G)$. Note
that usually even the well-definedness of $V_{\eta} \to{\rm L}^2(G)$ is
a non-trivial issue. The interest in isometries comes from the fact that
in this case there exists -- at least formally -- an inversion
formula, in the form of the weak operator integral
\[ \phi = \int_G V_{\eta} \phi (x) \pi(x) \eta d\mu_G(x) ~~.\]
A further pleasant feature of admissible vectors is that
the orthogonal projection onto the image $V_{\eta}({\cal H}_{\pi})
\subset {\rm L}^2(G)$ is given by convolution with 
$V_{\eta} \eta$, which entails that $V_{\eta}({\cal H}_{\pi})$ is a 
reproducing kernel Hilbert space. 
 
A vector $\eta$ is called {\bf weakly
admissible} if $V_{\eta}$ is a bounded one-to-one mapping
into ${\rm L}^2(G)$. Clearly, the condition that $V_{\eta}$
be one-to-one is equivalent to cyclicity of $\eta$.
 We call a representation {\bf weakly square-integrable}
if a weakly admissible vector exists, and {\bf strongly square-integrable}
if an admissible vector exists. It is well-known, that for irreducible
representations, the notions of weakly admissible and admissible
vectors coincide, and so do the weakly and strongly square-integrable
irreducible representations. But for the general case,
the two notions may well differ.

Both notions of square-integrability lead to subrepresentations
of the left regular representations: The wavelet transform
clearly intertwines $\pi$ with the left action of the group
(whichever function space on the group is under consideration). 
If $\eta$ is weakly admissible, then the unitary part of the 
polar decomposition of $V_{\eta}$ is a unitary equivalence
between $\pi$ and a subrepresentation of $\lambda_G$. Conversely,
Losert and Rindler proved \cite{LoRi}, that there exists
a cyclic vector $\eta$ for $\lambda_G$ iff $G$ is first countable.
Moreover they showed that the cyclic vector may be chosen 
continuous with compact support (in particular, in ${\rm L}^1(G)$),
which implies that it is in fact weakly admissible for $\lambda_G$,
by Young's inequality.
Hence $\lambda_G$ is weakly square-integrable iff $G$ is
first countable.
In order to obtain a weakly admissible vector for some 
invariant subspace, we only have to orthogonally project the weakly 
admissible vector for $\lambda_G$ into the subspace.
Thus we see that for first countable groups (and only for these)
weakly square-integrable representations are precisely those
which are equivalent to some subrepresentation of $\lambda_G$.

By contrast, strong square-integrability
is more complicated to check, as it will be seen to depend
on the modular function of the group: If $G$ is non-unimodular with
type-I regular representation, the notions
strongly and weakly square-integrable coincide, whereas the only
unimodular groups for which this is true are the discrete ones.
In the unimodular case, the existence of admissible vectors
can be characterized by a finite Plancherel measure condition
(at least when $\lambda_G$ is type-I).

All throughout the paper, $G$ denotes a (separable)
locally compact group
and $\upi$ a (strongly continuous, unitary) representation of $G$
on a separable Hilbert space ${\cal H}_{\pi}$.
$\lpraum{2}{G}$ denotes the usual ${\rm L}^2$-space on $G$
with respect to left Haar measure $\mu_G$. The left regular
representation $\lambda_G$ acts on $\lpraum{2}{G}$ by left translations.
For a function $f$ on $G$ let $f^* (x) := \overline{ f(x^{-1}) }$.

\section{Admissibility conditions for quasiregular representations}

In this section we consider a well-studied setting, where
$G  = \rhk{} \semdir H$
is a semidirect product of a vector group with a matrix group,
and $\pi$ is the quasiregular representation of that group acting
on $\lzw{}$. Various papers have been concerned with constructing
irreducible square integrable representations in such a setting
\cite{GrMoPa,Mu,Bo,BeTa,Fu} or with wavelet transforms obtained
from reducible representations \cite{IsKl,KlSt,WW}.

Throughout this section $G = \rhk{} \semdir H$ is a semidirect product,
with $H < {\rm GL}(k,\RR)$ a closed subgroup.
Elements of $G$ are denoted by $(x,h)$ with $x \in \rhk{}$ and
$h \in H$; the group law is then given by $(x_0,h_0)(x_1,h_1) =
(x_0+h_0x_1,h_0h_1)$. The modular function of $G$ can be written as 
$\Delta_G (x,h) = \Delta_H (h) |\det (h)|^{-1}$, and a left
Haar measure of $G$ is given by $d\mu_G(x,h) = | \det(h) |^{-1} dx d\mu_H(h)$.
For simplicity we will sometimes write $\Delta_G(h)$ instead of
$\Delta_G(0,h)$.
The quasiregular representation $\pi$ of $G$ acts on $\lzw{}$ by 
\[ (\pi (x,h) f) (y) = |\det (h)|^{-1/2} f(h^{-1}(y-x)) .\]

The closedness of $H$ in ${\rm GL}(n,\RR)$ may seem a
somewhat arbitrary condition (Lie subgroups might also work), 
but it is in fact not a real restriction, because of the following:
\begin{prop}
 Let $H$ be a subgroup of ${\rm GL}(n,\RR)$, endowed with some
 locally compact group topology. Assume that the semidirect product
 $\RR^k \semdir H$ is a topological semidirect product, and that
 the quasiregular representation has a nontrivial subrepresentation
 with an admissible vector. Then $H$ is a closed subgroup of
 ${\rm GL}(n,\RR)$, and the topology on $H$ is the relative topology.
\end{prop}
 
\begin{prf}
 Confer \cite[Proposition 5]{Fu2} for a detailed argument.
 In fact, the only property needed of $\pi$ is that it 
 has a non-trivial matrix coefficient vanishing at infinity.
 This requirement is met because $\pi$ has a common subrepresentation
 with the regular representation, and matrix coefficients
 of the latter are $C_0$-functions.
\end{prf}

The dual group $\rhkdu$ is the character group of $\rhk$,
suitably identified with the space of row vectors.
We define the Fourier transform as a mapping ${\cal F}:{\rm L}^2(\rhk)
 \to {\rm L}^2(\rhkdu)$ by letting
\[ {\cal F}(f) = \widehat{f}(\omega) = (2 \pi)^{-k/2}
 \int_{\rhk} f(x) e^{-i \omega x} dx ~~,\]
on ${\rm L}^1(\rhk) \cap {\rm L}^2(\rhk)$. Consequently
the Plancherel formula is given by
$\langle f, g \rangle = \langle \widehat{f}, \widehat{g} 
 \rangle$. 
Subrepresentations of the quasiregular representation are best 
described in terms of the dual action of $H$ on $\rhkdu$, which 
arises by duality from the action on $\rhk$. 
Under our identification of $\rhkdu$ with row vectors, the dual 
action is matrix multiplication on the right. Let $\rhkdu /H$ denote
the orbit space of this action, and let $q : \rhkdu \to \rhkdu /H$ be
the quotient map. For $\gamma \in \rhkdu$, we let $H_{\gamma}$
denote the stabilizer of $\gamma$ in $H$; it is a closed subgroup
of $H$.
For the discussion of subrepresentations of $\pi$, it is useful
to introduce the representation $\widehat{\pi}$ obtained by conjugating
$\pi$ with the Fourier transform on $\RR^k$. It is readily seen to 
operate on ${\rm L}^2(\widehat{\RR^k})$ via
 \begin{equation} \label{dual_rep}
 (\widehat{\pi} (x,h) \widehat{f}) (\gamma) = | \det (h) |^{1/2} e^{i \gamma
 \cdot x}
 f (\gamma h) ~~.
 \end{equation}
The action of $\widehat{\pi}$ allows to identify
subrepresentations in a simple way: Every invariant closed subspace
${\cal H} \subset {\rm L}^2(\RR^k)$ is of the form
\[ {\cal H} = {\cal H}_U = \{ g \in {\rm L}^2(\RR^k) : \widehat{g}
 \mbox{ vanishes outside of $U$ } \} ~~,\]
 where $U \subset \rhkdu$ is a measurable, $H$-invariant subset
(see \cite{Fu} for a detailed argument).
We let $\pi_U$ denote the subrepresentation acting on ${\cal H}_U$.
In view of the correspondence between subrepresentations
and invariant subsets of the dual, the dual orbit space becomes
a natural object of study.

\begin{remnt}{\bf Dual Orbit Space} 
\label{meas_ass}
 The structure of the dual orbit space is not only important 
 for the decomposition of the quasi-regular representation and
 for the construction of admissible vectors, but
 also for the decomposition of the regular representation of $G$,
 i.e., the Plancherel decomposition. For our discussion,
 the following two sets will be central
 \[ \Omega_c = \{ \omega \in \rhkdu : H_{\omega} \mbox{ is compact }
 \}~~,~~ \Omega_{rc} = \{ \omega \in \Omega_c : \omega H \mbox{ is
 locally closed } \} ~~.\]
 The set $\Omega_{rc} \subset \Omega_c$ consists of the ``regular'' orbits 
 in $\Omega_c$; i.e., it is the ``well-behaved'' part of $\Omega_c$.
 Loosely speaking, $\Omega_c$ is the set we have to deal with,
 and $\Omega_{rc}$ is the set we can deal with. Put more precisely:
 While Theorem \ref{adm_crit_elem} below shows that subrepresentations
 with admissible vectors necessarily correspond to invariant
 subsets $U$ of $\Omega_c$, the existence result in Theorem
 \ref{ex_ad_vec_qr} only considers subsets of the smaller set $\Omega_{rc}$.
 However, this distinction is not just a technical restriction
 inherent to our approach: As Example \ref{c_ex_Omega_c} below shows,
 subsets of $\Omega_c$ might not allow admissible vectors for the
 corresponding subrepresentations.

 Let us now collect some measure-theoretic properties of the two sets.
 $\Omega_c$ can be shown to be measurable;
 we have included a proof of that statement in the appendix. But
 usually $\Omega_c$ is not open, even when it is conull, as is
 illustrated by the example of ${\rm SL}(2,\ZZ)$: It is easy to see that
 $\Omega_c$ consists of all the vectors $(\omega_1,\omega_2)$ such
 that $\omega_1/\omega_2$ is irrational. This is a conull set with
 dense complement in $\widehat{\RR^2}$.
 
 By contrast, $\Omega_{rc}$ is always open (cf. the appendix
 for a proof). A pleasant consequence of this is that Glimm's
 Theorem \cite{Gl} applies (since $\Omega_{rc}$ is locally compact), which  
 entails a number of useful properties of the orbit space
 $\Omega_{rc} / H$: It is a standard Borel space having a 
 measurable cross section $\Omega_{rc} / H \to \Omega_{rc}$,
 and there exists a measurable transversal, i.e., a Borel
 subset $A \subset \Omega_{rc}$ meeting each orbit in precisely
 one point.

 Unfortunately, the example of ${\rm SL}(2,\ZZ)$ shows that $\Omega_{rc}$
 can be empty even when $\Omega_c$ is conull: Since the complement
 of $\Omega_c$ is dense, $\Omega_c$ contains no nonempty open set.
\end{remnt}

Let us now derive the admissibility condition for the quasiregular
representation, or more generally for subrepresentations.
As a matter of fact, the admissibility condition
can be given without any smoothness assumption. It is only when
we address the existence of functions fulfilling
the condition that we are forced to use more involved measure-theoretic
arguments. The theorem was derived for certain concrete
groups $H$ in \cite{IsKl,KlSt,MaZh};
the general version given here appears also in \cite{LWWW}.
Note that the admissibility condition also figures as a part
of the definition of the notion of ``projection generating
function'' in \cite[Definition 2.1]{GKT}. Thus the following
theorem also answers a question raised in \cite[Remark 2.6(b)]{GKT}:
There the authors observe that taking a projection generating function 
as wavelet gives rise to orthogonality relations among the 
wavelet coefficients which closely
resemble those for irreducible square-integrable representations,
even though the representation at hand is not irreducible.
This is readily explained by the isometry property guaranteed
by the admissibility condition. That these orthogonality relations also
arise in the non-irreducible setting is due to the fact that the
representation can be identified with a subrepresentation
of the Plancherel decomposition of the group. This is precisely the topic
of our paper. 

\begin{thm}
 \label{adm_crit_elem}
 Let $(\pi_U,{\cal H}_{U})$ be a subrepresentation of $\pi$
 corresponding to some invariant measurable subset $U$. Then
 \begin{eqnarray*}
 g \in {\cal H}_U \mbox{ is weakly admissible } & \Leftrightarrow &
  \mbox{the mapping } \gamma \mapsto \int_{H} |\widehat{g} (\gamma h)|^2 d\mu_H (h)  
 \mbox{ is positive} \\ & & \mbox{ and essentially bounded on $U$ } ~~,\\   
  g \in {\cal H}_U \mbox{ is admissible } & \Leftrightarrow & 
  \int_{H} |\widehat{g} (\gamma h)|^2 d\mu_H (h) = 1 \hspace{3mm} 
 (\mbox{ for } \mbox{ almost every } \gamma  \in U ) ~~.
 \end{eqnarray*}
 In particular, if $\pi_U$ has a weakly admissible vector, then
 $U \subset \Omega_c$ (up to a null set).
\end{thm}
\begin{prf}
 We start by explicitly calculating the ${\rm L}^2$-norm of $V_{g} f$,
 for $f,g \in {\cal H}_U$. The following computations are standard,
 see also \cite{BeTa,Fu,WW}; we include them for convenience.
\begin{eqnarray*}
 \left\| V_{g} f \right\|^2_{\lpraum{2}{G}}  & = & 
 \int_G \left| \langle f, \pi (x,h) g \rangle \right|^2 d\mu_G(x,h)
 \\
 & = &
 \int_G \left| \langle \widehat{f}, (\pi(x,h)g)^{\wedge} \rangle \right|^2 
       d\mu_G(x,h) \\
 & = & \int_G \left| \int_{\rhkdu{}} \widehat{f}(\gamma)
        |\det(h)|^{1/2}e^{-2\pi i \gamma x}
       \overline{\widehat{g}} (\gamma h) d\lambda(\gamma) \right|^2 d\mu_G(x,h)
        \\
 & = &
 \int_H \int_{\rhk{}} \left| \int_{\rhkdu{}} \widehat{f}(\gamma)
        e^{-2\pi i \gamma x} \overline{\widehat{g}} (\gamma h)
       d\lambda(\gamma)
        \right|^2 
        d\lambda(x) d\mu_H(h) \\
 & = & \int_H \int_{\rhk{}} \left| 
       {\cal F} (\phi_h) (x) \right|^2 d\lambda(x) d\mu_H(h). 
\end{eqnarray*}
Here $\phi_h(\gamma) = \widehat{f}(\gamma) \overline{\widehat{g}}(\gamma h)$,
and ${\cal F}$ denotes the Fourier transform on $\lpraum{1}{\rhkdu{}}$.
An application of Plancherel's formula to the last expression yields
\begin{eqnarray*}
 \int_H \int_{\rhkdu{}} \left| \phi_h(\gamma) \right|^2 d\lambda(\gamma)
      d\mu_H(h) 
& = &  \int_H \int_{\rhkdu{}} \left| \widehat{f}(\gamma) \right|^2
      \left| \widehat{g}(\gamma h) \right|^2 d\lambda(\gamma) d\mu_H(h) \\
& = & \int_{\rhkdu{}} \left| \widehat{f} (\gamma) \right|^2 \left( 
      \int_H \left| \widehat{g} (\gamma h) \right|^2 d\mu_H(h) \right) 
      d\lambda(\gamma)~.
 \end{eqnarray*}
 Now the mapping
 \[ \gamma \mapsto  \int_H \left| \widehat{g} (\gamma h) \right|^2 d\mu_H(h)
 \]
 is constant on orbits, due to left-invariance of $\mu_H$,
 and the admissibility criterion is proved.
 It is easily seen that whenever the stabilizer $H_{\gamma}$ is
 noncompact, we have
 \[  \int_H | \widehat{g} (\gamma h)|^2 d\mu_H(h) \in \{ 0, \infty \},
 \]
 (cf. also the proof of \cite[Theorem 10]{Fu}),
 hence $V_{g} f \in \lpraum{2}{G}$ entails that the pointwise product
 $\widehat{f} \widehat{g}$ vanishes a.e. outside of $\Omega_c$.
 In particular, a weakly
 admissible vector vanishes almost everywhere outside of $\Omega_c$, hence
 we obtain in such a case that $U \subset \Omega_c$ (up to a null set).
\end{prf}

For the construction of admissible vectors we need to
decompose Lebesgue-measure $\lambda$ on $\Omega_{rc}$ into certain
measures on the orbits and a measure on $\Omega_{rc}/H$.
Then we address the relationship of the measures on the orbits
to the Haar measure of $H$.

\begin{lemma}
\label{lem_meas_disint}
 \begin{itemize}
 \item[(a)] There exists a measure $\overline{\lambda}$ on $\Omega_{rc}/H$
 and on each orbit $\gamma H$ a measure $\beta_{\gamma H}$ such that for every
 measurable $A \subset \Omega_{rc}$ the mapping 
 \[ \gamma H \mapsto  \int_{\gamma H} \chi_A (\omega)
   d\beta_{\gamma H} (\omega) \]
 is $\overline{\lambda}$-measurable, and in addition
\[ \lambda (A) = \int_{\rhkdu/H} \int_{\gamma H} \chi_A (\omega)
   d\beta_{\gamma H} (\omega)
    d\overline{\lambda} (\gamma H) .\]
 \item[(b)] Let $(\overline{\lambda}, (\beta_{\gamma H})_{\gamma H
 \in \Omega_{rc}/H})$ be as in (a).
 For $\gamma \in \Omega_{rc}$ define $\mu_{\gamma H}$ as the image measure
 of $\mu_H$ under the projection map $p_{\gamma} :
 h \mapsto \gamma h$. $\mu_{\gamma H}$
 is a $\sigma$-finite measure, and its definition is
 independent of the choice of representative $\gamma$. Then,
 for almost all $\gamma \in \Omega_{rc}$, the
 $\mu_{\gamma H}$ and $\beta_{\gamma H}$
 are equivalent, with globally Lebesgue-measurable Radon-Nikodym-derivatives:
 There exists an (essentially unique)
 Lebesgue-measurable function $\kappa: \Omega_{rc} \to \RR^+$ such that for
 $\omega \in \gamma H$,  
 \[ \frac{d\beta_{\gamma H}}{d\mu_{\gamma H}} (\omega) = \kappa(\omega) ~~.\]
 \item[(c)] 
 The function $\kappa$ fulfills the {\bf semi-invariance relation}
 \begin{equation} \label{sem_inv_konk}
 \kappa(\omega h) = \kappa (\omega) \Delta_G(h)^{-1}
 ~~.\end{equation}
 In particular, $\kappa$ is $H$-invariant iff $G$ is unimodular. 
 In that case, we can in fact assume that $\kappa=1$ almost everywhere.
 This choice determines the measure $\overline{\lambda}$ uniquely.
 \end{itemize}
\end{lemma}
\begin{prf}
 Statement $(a)$ is a classical result from measure theory;
 see for instance \cite[Theorem 2.1]{KlLi}.
 The standardness of $\Omega_{rc}/H$ is decisive.
 
 In order to prove part $(b)$, well-definedness and $\sigma$-finiteness
 of $\mu_{\gamma H}$ follow from compactness of $H_{\gamma}$. The
 independence of the representative $\gamma$ of the orbit follows
 from the fact that $\mu_H$ is left-invariant, and the dual action
 is on the right. To compute the Radon-Nikodym derivative $\kappa$,
 we first introduce an auxiliary function
 $\ell : \Omega_{rc} \to \RR_0^+$: Fix a Borel-measurable transversal
 $A \subset \Omega_{rc}$ of the $H$-orbits.
 Then the mapping $\tau : A \times H \to \Omega_{rc}$, $\tau(\omega,h)
 = \omega h$ is bijective and continuous, hence, since $A \times H$
 is a standard Borel space, $\tau^{-1}: \Omega_{rc} \to A \times H$
 is Borel as well, by \cite[Theorem 3.3.2]{Ar}.
 If we let $\tau^{-1}(\gamma)_H$ denote the $H$-valued coordinate
 of $\tau^{-1}(\gamma)$, then $\ell(\gamma) :=
 \Delta_G(\tau^{-1}(\gamma)_H)$ is a Borel-measurable mapping.
 Since $\Delta_G$ is constant on every compact subgroup (in
 particular on all the little fixed groups of elements in $\Omega_{rc}$),
 a straightforward calculation shows that $\ell$ satisfies the
 semi-invariance relation $\ell(\omega h) = \ell (\omega) \Delta_G(h)^{-1}$.

 Next fix an orbit $\gamma_H$ and let us compare the measures
 $\beta_{\gamma H}$ and $\ell \mu_{\gamma H}$: Since
 \[ d\mu_{\gamma H}(\omega h) = \Delta_H(h) d\mu_{\gamma H}(\omega)
  ~~\mbox{ and }~~
 d\beta_{\gamma H}(\omega h) = | \det(h) |  d\beta_{\gamma H}(\omega) ~~,\]
 the definition of $\ell$ ensures that 
 $\ell \mu_{\gamma H}$ and $\beta_{\gamma H}$ behave
 identically under the action of $H$. Moreover,
 they are $\sigma$-finite and quasi-invariant, hence equivalent.
 Since they have the same behaviour under translations by $H$,
 the Radon-Nikodym derivative turns out to be a positive constant on the
 orbit.
 Summarizing, we find for $\omega \in \gamma H$ that 
 \[ \frac{d \beta_{\gamma H}}{d \mu_{\gamma H}} (\omega) = 
 \ell(\omega) c_{\gamma H} ~~,\]
 with $\ell, c_{\gamma H} >0$, 
 and it remains to show that $c_{\gamma H}$ depends measurably on the orbit. 

 For this purpose pick a relatively compact open neighborhood $B \subset H$
 of the identity. Then $AB = \tau(A \times B) \subset \Omega_{rc}$
 is Borel-measurable, as a continuous image of a standard space,
 hence $\chi_{AB}$, the indicator function of $AB$, is
 a Borel-measurable function. Both
 \[ \phi_1 : \gamma H \mapsto \int_{\gamma H} \chi_{AB} (\omega)
 d\beta_{\gamma H}(\omega) \]
 and 
 \[ \phi_2 : \gamma H \mapsto \int_{\gamma H} \chi_{AB} (\omega) \ell(\omega) 
 d\mu_{\gamma H}(\omega)
 \]
 are measurable functions: The first one is by choice of the
 $\beta_{\gamma H}$, see part (a). The second one is measurable
 by Fubini's theorem, applied to the mapping $(\omega,h) \mapsto
 \chi_{AB}(\omega h) \ell(\omega h)$ on $\rhkdu \times H$
 (recall the definition of $\mu_{\gamma H}$).

 In addition, both functions are finite and positive on $\Omega_{rc}$.
 We have 
\[ \phi_2(\gamma H) = \int_{\gamma H} \chi_{AB} (\omega) \ell(\omega) 
 d\mu_{\gamma H}(\omega) = \int_{p_{\gamma}^{-1}(AB)} \Delta_G(h) d\mu_H(h)~,
\]
 and $p_{\gamma}^{-1}(AB)$ is relatively compact and open, hence it
 has finite and positive Haar measure.
 Since in addition $\Delta_G$ is positive and bounded on
 $p_{\gamma}^{-1}(AB)$, we find $0 < \phi_2(\gamma H) < \infty$. 
 Hence 
 \[\phi_1(\gamma H) = c_{\gamma H} \phi_2(\gamma H)
 \]
 can be solved for $c_{\gamma H}$, which thus turns out to depend
 measurably upon $\gamma H$. Hence 
 \[ \kappa(\omega) = \frac{d \beta_{\gamma H}}{d \mu_{\gamma H}} =
 \ell(\omega) c_{\gamma H} \]
 is a Lebesgue-measurable function.

 The remaining part $(c)$ is simple to prove: The semi-invariance
 relation of $\ell$ entails the relation for $\kappa$. The normalization
 is easily obtained: If $\kappa$ is constant on the orbits,
 it defines a measurable mapping $\overline{\kappa}$ on
 $\Omega_{rc}$. If we replace each $\beta_{\gamma H}$ by
 $\mu_{\gamma H}$, we can make up for it by taking
 $\overline{\kappa}(\gamma H) d\overline{\lambda}(\gamma H)$ as the new
 measure on the orbit space. The new choice has the desired properties. 
 The uniqueness of $\overline{\lambda}$ follows from the usual
 Radon-Nikodym arguments.
\end{prf}

\begin{rem}
 Let us for the rest of the paper fix a choice
 of $\overline{\lambda}$. Note that this also
 uniquely determines the function $\kappa$.
 In the unimodular case we take $\kappa$ to be $1$, which
 in turn determines $\overline{\lambda}$ uniquely. 

 As we shall later see,
 the choice of a pair $(\overline{\lambda},\kappa)$ corresponds exactly
 to a choice of Plancherel measure and the associated family of
 Duflo-Moore operators (at least on a subset of $\widehat{G}$).
\end{rem}

Before we turn to the construction of admissible vectors,
we introduce some notation to help clarify the construction:
To a function $\widehat{g}$ on $U$ we associate two auxiliary
$H$-invariant functions $T_H (\widehat{g})$ and  $S_H (\widehat{g})$
such that admissibility of $g$ translates to a condition on 
$T_H (\widehat{g})$ and square-integrability to a 
condition on $S_H (\widehat{g})$.
\begin{defn}
For a measurable function $\widehat{g}$ on $\Omega_{rc}$, 
let $T_H (\widehat{g})$ denote the function
\[ T_H(\widehat{g}) (\omega) := \left( \int_{\omega H}
 | \widehat{g}(\gamma) |^2
 d\mu_{\omega H}(\gamma) \right)^{1/2} = 
 \left( \int_{\omega H}
 | \kappa(\omega)^{-1/2} \widehat{g}(\gamma) |^2
 d\beta_{\omega H}(\gamma) \right)^{1/2}
 ~~.\]
$T_H(\widehat{g})$ is a measurable, $H$-invariant mapping
$\Omega_{rc} \to \RR_0^+ \cup \{ \infty \}$. The admissibility condition can
then be reformulated:
\begin{equation}
\label{adm_T_H}
 g \in {\rm L}^2(U) \mbox{ is admissible }
\Leftrightarrow T_H(\widehat{g}) \equiv 1 
~~(\mbox{ a.e. on U}) ~~.\end{equation} 
Similarly, weak admissibility is equivalent to the requirement
that $T_H(\widehat{g}) \in {\rm L}^{\infty}(U)$ and $T_H(\widehat{g})>0$
almost everywhere.
We can also define
\[
 S_H(\widehat{g}) (\omega) := \left( \int_{\omega H} | \widehat{g}(\gamma) |^2
 d\beta_{\omega H}(\gamma) \right)^{1/2} ~~.
\]
By our choice of measures, $S_H$ and $T_H$ coincide iff $G$ is unimodular.
Both $T_H(\widehat{g})$ and $S_H(\widehat{g})$ may (and will) be
regarded as functions on the quotient space $U/H$. 
By the choice of the $\beta_{\omega H}$, 
\begin{equation}
\label{L2_S_H}
 \int_U |\widehat{g}(\omega) |^2 d\omega = \int_{U/H} | S_H(\widehat{g})
(\omega H) |^2 d\overline{\lambda} (\omega H) ~~,
\end{equation}
so that $\widehat{g}$ is square-integrable iff $S_H(\widehat{g})$ is
a square-integrable function on $U/H$. 
\end{defn}

Now we can address the existence of admissible vectors.
The following theorem is essentially the same as \cite[Theorem 1.8]{LWWW}.

\begin{thm}
 \label{ex_ad_vec_qr}
 Let $U \subset \Omega_{rc}$ be measurable and $H$-invariant.
 Then $\pi_U$ has a weakly admissible vector. It has an
 admissible vector iff either
 \begin{itemize}
 \item[(i)] $G$ is unimodular and $\overline{\lambda}(U/H) < \infty$. 
 \item[(ii)] $G$ is non-unimodular.
 \end{itemize}
\end{thm}

\begin{prf}
 Recall that by the last remark we have for each admissible 
 vector $g$ that $T_H ({\widehat{g}})$ is constant almost everywhere.
 At the same time, in the unimodular case it is square-integrable as
 a function on $U/H$, because of $S_H = T_H$.
 This shows the necessity of $(i)$ in the unimodular case.

 To prove the existence of admissible vectors, we first construct
 a function $\widehat{g}$ on $U$ fulfilling the admissibility
 condition (\ref{adm_T_H}), and then modify the construction
 to provide for square-integrability. 

 For this purpose we recycle the sets $A \subset \Omega_{rc}$
 and $B \subset H$ from the proof of Lemma \ref{lem_meas_disint}.
 We already observed there that $\widehat{f} = \chi_{AB}$ is
 Lebesgue-measurable, and that $T_H(\widehat{f})$ is positive and
 finite almost everywhere on $U$. Hence we may define
 $\widehat{g} = \widehat{f} / T_H(\widehat{f})$, which fulfills
 the admissibility criterion. In the unimodular case, with
 equation (\ref{L2_S_H}) together with $S_H = T_H$ shows that
 $\widehat{g} \in {\rm L}^2(U)$.

 In the non-unimodular case, we modify $g$ as follows:
 For every $\gamma \in U$, the compactness
 of $p_{\gamma}^{-1}(AB)$ entails that $\Delta_G$ is bounded
 on that set; hence $C$ is bounded on $AB$. Then
 $S_H(\widehat{g})$ is positive and finite almost
 everywhere. Since $\overline{\lambda}$ is $\sigma$-finite,
 we can write $U/H = \bigcup_{n \in \NN} V_n$, with disjoint $V_n$
 of finite measure, such that 
 in addition $S_H(\widehat{g})$ is bounded on each $U_n$
 (here we regard $S_H(\widehat{g})$ as a function on the quotient).
 In particular, $S_H(\widehat{g}) \cdot \chi_{V_n}$ is a square-integrable
 on $U/H$.
 Now let $U_n \subset U$ be the inverse image of $V_n$ under the quotient
 map, and for $h_0 \in H$ and $n, k_n \in \NN$, denote by
 \[ 
  \widehat{g_n}(\omega)
  :=  \Delta_H(h_0)^{k_n/2} \widehat{f_2} (\omega h_0^{k_n}) \cdot
 \chi_{U_n} (\omega) ~~.
 \]
 Then the normalization ensures that $\widehat{g_n}$ has the
 following properties:
 \begin{equation}
 \label{part_adm}
  T_H({\widehat{g_n}}) = \chi_{U_n} 
 \end{equation}
 and 
 \begin{equation}
 \label{norm_g_n}
 S_H(\widehat{g_n}) = \Delta_H(h_0)^{k_n/2}
 | \det (h_0) |^{-k_n/2} S_H(\widehat{g}) \cdot \chi_{U_n}
 = \Delta_G(h_0)^{k_n/2} S_H(\widehat{g}) \cdot \chi_{U_n}
  ~~.
 \end{equation}
 Hence the following construction gives an admissible vector: Choose
 $h_0 \in H$ such that $\Delta_G(h_0)<1/2$, pick
 $k_n \in \NN$ satisfying
 \begin{equation}
 \label{def_k_n}
  2^{-k_n} \| S_H(\widehat{g}) \cdot
 \chi_{U_n} \|_2^2 < 2^{-n} 
 \end{equation}
 and let 
 $\widehat{\tilde{g}}(\omega) := \delta_H(h_0)^{k_n/2} \widehat{f_2}
 (\omega h_0^{k_n})$, for $\omega \in U_n$. Then (\ref{part_adm}) implies
 that $T_H(\widehat{\tilde{g}}) = 1$ a.e., whereas (\ref{norm_g_n})
 and (\ref{def_k_n}) ensure that $S_H(\widehat{\tilde{g}}) \in
 {\rm L}^2(U/H,\overline{\lambda})$.

 A weakly admissible vector for $\pi_U$ (which is missing in the
 unimodular case) can be obtained by similar (somewhat simpler) methods.
\end{prf}

\begin{rem}
\label{c_ex_Omega_c}
 In Theorem \ref{adm_crit_elem} we cannot replace $\Omega_{rc}$ by
 the bigger set $\Omega_c$. To give a non-unimodular example, let $H = \{
 2^k h : k \in \ZZ , h \in {\rm SL}(2,\ZZ) \}$,
 which is a discrete subgroup of ${\rm GL}(2,\RR)$.
 Whenever $(\gamma_1,\gamma_2) \in \rhkdu$ is such that $\gamma_1 /\gamma_2$
 is irrational, the stabilizer of $(\gamma_1,\gamma_2)$ in $H$
 has two elements. Hence the set $\Omega_{c}$ is a conull subset in $\rhkdu$,
 whereas (as we already noted) $\Omega_{rc}$ is empty.
 $H$ operates ergodically on $\rhkdu$ (already ${\rm SL}(2,\ZZ)$ does,
 \cite[2.2.9]{Zi}), and hence $\pi$ is an irreducible representation.
 But it has been shown that for discrete dilation groups
 $\pi$ is never irreducible and square-integrable, see \cite[Remark 12]{Fu}.
\end{rem}



Let us now give a short summary of the steps which have to be
carried out for the construction of wavelet transforms from semidirect
products: 
\begin{enumerate}
 \item Compute the $H$-orbits in $\rhkdu{}$, possibly by giving a
  parametrization of $\rhkdu /H$.
 \item Determine the set $\Omega_{rc}$. If $\lambda (\Omega_{rc}) = 0$, stop.
 \item Parametrize each orbit in $\Omega_{rc}$ and determine the image
  $\mu_{\gamma H}$ of Haar measure under the projection map
  $h \mapsto \gamma h$. 
 \item Compute the measure decomposition $d\lambda(\gamma) =
  d\beta_{\omega H} (\gamma) d\overline{\lambda} (\omega H)$.
 \item Compute the Radon-Nikodym derivative $C$.
 \item The admissibility condition can then be formulated for
  subsets of $\Omega_{rc}$ just as in Theorem \ref{adm_crit_elem}.
  Theorem \ref{ex_ad_vec_qr} ensures the existence of admissible
  vectors. 
\end{enumerate}

Since the final step -- the actual construction of admissible vectors --
is missing, the description is somewhat incomplete. Clearly
the construction given in the proof of Theorem \ref{ex_ad_vec_qr}
is not very practical, but it seems doubtful to us that a
more explicit method is available which works in full generality.
In many cases where parametrizations
of orbits and orbit spaces are possible, they can be given 
differentiably. Then computing the various measures and Radon-Nikodym 
derivatives reduces to computing the Jacobians of those parametrizations. 
We expect that in such a setting the construction 
of admissible vectors should also be facilitated.

For the remainder of this section we want to focus on
the case that $G$ is unimodular. The main motivation for the following
proposition is to show that certain subrepresentations of $\pi$
do not have admissible vectors. In the light of Theorem \ref{ex_ad_vec_qr},
this amounts to proving that $\overline{\lambda}(U/H)$ is infinite,
for the $H$-invariant set $U \subset \Omega_{rc}$ under consideration.

The argument proving the proposition
employs the action of the scalars on the orbit space $\Omega_{rc}/H$.
The group of scalars could be replaced by any group 
$A \subset {\rm GL}(n,\RR)$ which normalizes $H$. Symmetry
arguments of this type might help simplify the calculation of
the various measures which come up.

$\RR^+$ operates on
$\rhkdu/H$ by multiplication, i.e., if $a \in \RR^+$ then
$a \cdot (\gamma H) = (a \gamma) H$ is an operation. Obviously
$\Omega_{rc}$ is invariant, so that
we obtain an operation on $\Omega_{rc}/H$. The next proposition
gives the behaviour of $\overline{\lambda}$ under this action. 

\begin{prop}
\label{symm_arg}
 Assume that $G$ is unimodular. 
 Let the measures $\overline{\lambda}$ and $\mu_{\gamma H}$ be as in Lemma 
 \ref{lem_meas_disint}.
 For $a \in \RR^+$ and $\gamma \in \rhkdu{}$
 let $a^* (\mu_{\gamma H})$ denote the image measure of $\mu_{\gamma H}$
 on $\gamma H a$, i.e., for measurable $B \subset \gamma H a$ let 
 $a^* (\mu_{\gamma H}) (B) := \mu_{\gamma H} (B a^{-1})$. Moreover let 
 the measure $\overline{\lambda}_a$ be given by $\overline{\lambda}_a (B)
 := \overline{\lambda} (B a)$ $(B \subset \rhkdu{} /H$ measurable$)$.
 Then on $\Omega_{rc}/H$ the following relations hold:
 \begin{eqnarray*}
    \mu_{a \gamma H} & = &  a^*(\mu_{\gamma H}),\\
 \overline{\lambda}_a & = & a^n \overline{\lambda}.
 \end{eqnarray*}
\end{prop}

\begin{prf}
 The first equality is immediate from the definitions of $\mu_{\gamma H}$
 and $\mu_{a \gamma H}$.
 For the second equation let us introduce the following notation:
 If $f: \Omega_{rc} \to \RR$ is a positive, measurable function, let $q(f)$
 denote the function on $\Omega_{rc}/H$ defined by
 \[ q(f) (\gamma H) := \int_{\gamma H} f(\omega) d\mu_{\gamma H} (\omega) .\]
 Moreover let $f_a (\omega) := f (\omega a^{-1})$, for all
 $\omega \in \Omega_{rc}$ and
 $a \in \RR^+$. From the first equation we obtain
 \begin{eqnarray*}
    q (f_a) (\gamma H) & = & \int_{\gamma H} f (\omega a^{-1}) d\mu_{\gamma H}
 (\omega) \\
 & = & \int_{\gamma H a^{-1}} f (\omega) d\mu_{\gamma H a^{-1}}
 (\omega) \\
 & = & q (f) (\gamma H a^{-1}).
 \end{eqnarray*}  
 Using this equation, we compute
 \begin{eqnarray*}
  \int_{\Omega_{rc}/H} q(f)(\gamma H) d\overline{\lambda} (\gamma H) & = &
  a^{-n} \int_{\Omega_{rc}} f_a (\omega) d\lambda (\omega) \\
 & = & a^{-n} \int_{\Omega_{rc}/H} q(f_a) (\gamma H) d\overline{\lambda}
 (\gamma H) \\
 & = & a^{-n} \int_{\Omega_{rc}/H} q(f) (\gamma H a^{-1})
 d\overline{\lambda} (\gamma H) \\
 & = & a^{-n} \int_{\Omega_{rc}/H} q(f) (\gamma H)
 d\overline{\lambda}_a (\gamma H)
 \end{eqnarray*}
 Using arguments similar to the one used in the proof of Theorem
 \ref{ex_ad_vec_qr}, it is readily seen that
 for each measurable $A \subset \Omega_{rc}/H$ there
 exists a positive measurable $f$ on $\Omega_{rc}$ with $q(f) = \chi_A$.
 Hence we have shown the second equation.
\end{prf}

As a first consequence we obtain that admissible vectors exist
only for proper subsets of $\Omega_{rc}$. This was already noted
(in the special case where $\Omega_{rc}$ is conull in $\rhkdu$) in
\cite{LWWW}, Theorem 1.8.

\begin{cor}
 Assume that $G$ is unimodular, and that
 $U := \Omega_{rc}$ is not a nullset. Then
 the subrepresentation $\pi_U$ is not strongly square integrable.
\end{cor}
\begin{prf}
 By assumption we have $\overline{\lambda} (\Omega_{rc}/H) > 0$, and we need
 to show that $\overline{\lambda} (\Omega_{rc}/H) = \infty$. But
 $a \Omega_{rc} = \Omega_{rc}$ and  Proposition \ref{symm_arg}
 yield $\overline{\lambda} (\Omega_{rc}/H) =
 \overline{\lambda} (a \cdot \Omega_{rc}/H) =
 |a|^{-k}  \overline{\lambda} (\Omega_{rc}/H)$.
\end{prf}

It is well known that, given a square integrable 
representation $\sigma$ of a locally compact group $G$, every
vector in ${\cal H}_{\sigma}$ is admissible iff $\sigma$ is irreducible
and $G$ is unimodular. Hence irreducible representations would be
particularly useful, having no restrictions at all on admissible vectors.
But the following corollary excludes irreducible representations from our 
setting. The statement was proved first in \cite{Fu_Diss} by a 
technique employing the Fell topology of the group.
\begin{cor}
 Let $G$ be unimodular. 
 Then the quasiregular representation $\pi$ does not contain any 
 irreducible square-integrable subrepresentations.
\end{cor}
\begin{prf}
 Assume the contrary and let $\pi_U$ be an irreducible square integrable
 subrepresentation. Here $U$ denotes the corresponding $H$-invariant subset
 of $\rhkdu$. Then, by \cite[Theorem 1.1]{Anh}, $U$ is (up to a null
 set) an orbit of positive measure, hence open (by Sard's Theorem). 
 In particular $U \subset \Omega_{rc}$, and $\overline{\lambda}(\{ U \})
 > 0$. 

 From the fact that ${\cal H}_U$ has admissible vectors we conclude
 that $\overline{\lambda} (\{U \}) < \infty$. On the other hand,
 an easy connectedness argument shows that
 for each $\gamma \in U$, the ray $\RR ^+ \gamma$ is contained in the open
 orbit $U$. Hence the same argument which proved the previous corollary
 shows that $\overline{\lambda} (\{ U \}) = \infty$, which yields the
 desired contradiction.
\end{prf}

\section{Plancherel measure and admissibility}

This section is devoted to a short review of Plancherel theory
and its relations to generalized wavelet transforms. For a
more detailed (yet still short) account of Plancherel theory, 
consult \cite[Section 7.5]{Fo}. 
Unfortunately, there does not seem to exist a widely accessible
exposition of Plancherel theory that covers non-unimodular groups 
in sufficient detail; a good source for the unimodular case is
 \cite{DiC}. For
the non-unimodular theory, the original papers \cite{KlLi,Ta,DuMo} 
are probably still the best sources (though rather technical at times).

 All throughout this section, $G$ is a second countable locally compact
 group.
 The starting point for the definition of the Plancherel transform is 
 the operator valued Fourier transform on ${\rm L}^1(G)$. Given
 $f \in {\rm L}^1(G)$ and $\sigma \in \widehat{G}$, we define
 \[ {\mathcal F}(f) (\sigma) := \sigma(f) := \int_G f(x) \sigma(x) d\mu_G(x) ~~,\]
 where the integral is taken in the weak operator sense. 
 As direct consequences of the definition we have $\| \sigma(f) \|_{\infty}
 \le \| f \|_1$ and $\sigma(f\ast g) = \sigma(f) \circ \sigma(g)$.

 The Plancherel transform is obtained by extending the Fourier transform
 from ${\rm L}^1(G) \cap {\rm L}^2(G)$ to ${\rm L}^2(G)$. 
 The non-unimodular part of the following Plancherel theorem is
 due to Duflo and Moore \cite[Theorem 5]{DuMo}, whereas the
 unimodular version may be found in \cite{DiC}.
\begin{thm}
\label{Pl-Thm}
 Let $G$ be a second countable locally compact group having a type-I
regular representation. Then there exists a measure $\nu_G$ on $\hat{G}$
and a measurable field $(K_{\sigma})_{\sigma \in \hat{G}}$ of selfadjoint
positive operators with densely defined inverse, with the following properties:
\begin{enumerate}
\item[(i)] For $f \in {\rm L}^1(G) \cap {\rm L}^2(G)$ and
 $\nu_G$-almost all $\sigma \in \hat{G}$, the closure of the operator
 $\sigma(f) K_{\sigma}^{1/2}$ is a Hilbert-Schmidt operator on
 ${\mathcal H}_{\pi}$. 
\item[(ii)] The map ${\rm L}^1(G) \cap {\rm L}^2(G) \ni f \mapsto
 ([{\sigma(f) K_{\sigma}^{1/2}}])_{\sigma \in \widehat{G}}$ extends
 to a unitary equivalence 
 \[ {\mathcal P}: {\rm L}^2(G) \to {\mathcal B}_2^{\oplus} :=
 \int^{\oplus}_{\hat{G}} {\mathcal B}_2({\mathcal H}_{\sigma}) d\nu_G(\sigma) ~~.\]
 This unitary operator is called the {\bf Plancherel transform} of $G$.
 It intertwines the two-sided representation $\lambda_G \times \rho_G$ with
 $\int_{\hat{G}}^{\oplus} \sigma \otimes \overline{\sigma} d\nu_G(\sigma)$.
\item[(iii)] $G$ is unimodular iff almost all $K_{\sigma}$ are scalar
 multiples of the identity operator. In this case we fix $K_{\sigma}
 = {\rm Id}_{{\mathcal H}_{\sigma}}$.
\item[(iv)] Once a measurable choice $(K_{\sigma})_{\sigma \in \widehat{G}}$
 of Duflo-Moore operators is made, the Plancherel measure
 $\nu_G$ is uniquely determined. This applies in particular to the unimodular
 case, where $K_{\sigma} = {\rm Id}_{{\mathcal H}_{\sigma}}$
 leads to a unique definition of $\nu_G$.
\end{enumerate}
\end{thm} 

 In the following $\widehat{f}$ denotes the Plancherel transform 
 of the ${\rm L}^2$-function $f$; in particular in the non-unimodular
 case it should not be confused with the Fourier transform.

 A further important feature of the Plancherel transform is the decomposition
 of intertwining operators: If $T : {\rm L}^2(G) \to {\rm L}^2(G)$ is a 
 bounded operator which commutes with left translations, then 
 there exists a measurable field of bounded operators $(T_{\sigma})_{\sigma
 \in \widehat{G}}$ with $\| T_{\sigma} \|_{\infty}$ uniformly bounded,
 such that 
 \[ T = \int_{\widehat{G}}^{\oplus} {\rm Id}_{{\mathcal H}_{\sigma}} \otimes
 T_{\sigma} d \nu_G (\sigma) ~~.\]
 This applies in particular to the projection onto invariant subspaces.
 The obvious analogue for the right action of $G$ holds as well.
 This decomposition property is the key feature from the point of
 view of generalized wavelet transforms. First of all it 
 provides systematic access to the subrepresentations of the regular
 representations, and by the discussion at the end of the Introduction,
 those subrepresentations exhaust all situations of interest.
 Moreover, for a given admissible vector $\eta$, there exist
 essentially two Plancherel transforms of interest: 
 The Plancherel transform of $\eta$ as ${\rm L}^2$-function, and
 the decomposition of the intertwining operator $V_{\eta}$. Relating
 those two objects enables us to formulate admissibility conditions
 and to construct vectors which fulfill them. The following
 theorem summarizes the results of \cite{Fu3}. Note that
 the admissibility conditions in part (b) of the theorem are
 only sufficient. Note also that \cite{Fu3} uses the operators
 $C_{\sigma} = K_{\sigma}^{-1/2}$.

\begin{thm}
\label{ex_adm_vec_abs}
 Assume that $\lambda_G$ is type-I. Let ${\cal H} \subset {\cal B}_2^{\oplus}$
 be an invariant subspace and denote by $P$ the projection onto ${\cal H}$.
 Then there exists a measurable family of projections
 $(P_{\sigma})_{\sigma \in \widehat{G}}$ such that
  \[  P = \int_{\hat{G}}^{\oplus} {\rm Id}_{{\mathcal H}_{\sigma}} 
  \otimes P_{\sigma} d\nu_G(\sigma) ~~.\]
 \begin{enumerate} 
 \item[(a)] Assume that $G$ is unimodular. Then $\eta \in {\cal H}$
 is 
 \begin{eqnarray*}
 \mbox{ weakly admissible } & \mbox{ iff } & \widehat{\eta} (\sigma) \mbox{ is
 injective on $\nu_G$-a.e. $P_{\sigma}({\cal H}_{\sigma})$,}
 \\ & & \mbox{ and }
 \sigma \mapsto \| \widehat{\eta}(\sigma) \|_{\infty}  
 \mbox{ is an ${\rm L}^{\infty}$-function.}
 \\
 \mbox{ admissible } & \mbox{ iff } &  \widehat{\eta} (\sigma)  
 \mbox{ is an isometry on $P_{\sigma}({\cal H}_{\sigma})$, for
 $\nu_G$-a.e. $\sigma$}. \\
 \mbox{ ${\cal H}_U$ has an admissible vector } & \mbox{ iff } &
  \int_{\widehat{G}} \dim (P_{\sigma} ({\cal H}_{\sigma})) d\nu_G(\sigma)
 < \infty
 \end{eqnarray*}
 $\lambda_G$ has an admissible vector iff $G$ is discrete.
 \item[(b)] Assume that $G$ is non-unimodular. Assume
 that $\eta \in {\cal H}$ is such that $\widehat{\eta}(\sigma)^*
 K_{\sigma}^{-1/2}$
 extends to a bounded operator on ${\cal H}_{\sigma}$, for $\nu_G$-almost
 every $\sigma \in \widehat{G}$.
 Then $\eta \in {\cal H}$ is 
 \begin{eqnarray*}
 \mbox{ weakly admissible } & \mbox{ if } & \widehat{\eta} (\sigma) \mbox{ is
 injective on $\nu_G$-a.e. $P_{\sigma}({\cal H}_{\sigma})$,}
 \\ & & \mbox{ and }
 \sigma \mapsto \| K_{\sigma}^{-1/2} \widehat{\eta}(\sigma) \|_{\infty}  
 \mbox{ is an ${\rm L}^{\infty}$-function.}
 \\
 \mbox{ admissible } & \mbox{ if } & K_{\sigma}^{-1/2} \widehat{\eta}(\sigma) 
 \mbox{ is an isometry on
  $P_{\sigma}({\cal H}_{\sigma})$, for $\nu_G$-a.e. $\sigma$ ~~.} 
  \end{eqnarray*}
  There exists a vector $\eta \in {\cal B}_2^{\oplus}$ fulfilling
  the admissibility condition for ${\cal H}= {\cal B}_2^{\oplus}$.
  Hence  $\lambda_G$ -- and thus every subrepresentation thereof --
  has an admissible vector.
 \end{enumerate}
\end{thm}

As a matter of fact, the statements concerning weak admissibility
cannot be found in \cite{Fu3}, but the arguments dealing with
admissibility can be easily modified to cover these results as well.
The statements somewhat simplify when we consider multiplicity-free
representations. Since the quasi-regular representation will turn
out to be multiplicity-free, we find it useful to work out this
particular case in some detail:

\begin{thm}
\label{adm_crit_mfree}
 Assume that $\lambda_G$ is type-I.
 Let $\pi$ be a multiplicity-free weakly square-integrable representation
 of $G$. Then there exists a $\nu_G$ measurable set $\Sigma \subset
 \widehat{G}$ such that  
 \[ \pi \simeq \int_{\Sigma} \sigma d\nu_G(\sigma) ~~.\]
  For the following, we assume that $\pi$ is in fact realized as
  the direct integral.
  \begin{enumerate}
  \item[(a)] Assume that $G$ is unimodular. $\eta = (\eta_{\sigma})_{\sigma
  \in \Sigma} \in {\cal H}_{\pi}$ is admissible iff
  $\| \eta_{\sigma} \| = 1$ for $\nu_G$-every
  $\sigma \in \Sigma$. $\pi$ is strongly square-integrable iff $\nu_G(\sigma)
  < \infty$.
  \item[(b)] Assume that $G$ is non-unimodular. Let $\eta =
  (\eta_{\sigma})_{\sigma \in \Sigma}$ be such that for $\nu_G$-a.e.
 $\sigma \in \Sigma$, 
  $\eta_{\sigma} \in {\rm dom} (K_{\sigma}^{-1/2})$, with
 $\| K_{\sigma}^{-1/2} \eta_{\sigma}
  \| = 1$. Then $\eta$ is admissible. There exist vectors $\eta \in 
  {\cal H}_{\pi}$ fulfilling this admissibility condition.
  \end{enumerate} 
\end{thm}

A consequence of the theorem is the following proposition,
which shows that Plancherel measure is in fact characterized
by the admissibility condition. This observation will allow
us to identify the quotient measure $\overline{\lambda}$
obtained in Section 1 with Plancherel measure.
\begin{prop}
\label{adm_and_pm}
 Assume that $\lambda_G$ is type-I. Let $(K_{\sigma})_{\sigma \in
 \widehat{G}}$ be a measurable choice of Duflo-Moore operators
 -- for $G$ unimodular, $K_{\sigma} = {\rm Id}$ -- and let 
 $\nu_G$ be the corresponding Plancherel measure. 
 Let $\pi, \Sigma$ be as in Theorem \ref{adm_crit_mfree}, and
 assume in addition that there exists an admissible vector for $\pi$;
 i.e., assume in the unimodular case that $\nu_G(\Sigma) < \infty$ .
 Let $\tilde{\nu}$ be a Borel measure on $\Sigma$ which
 is equivalent to $\nu_G$, and consider the representation
 \[ \tilde{\pi} = \int_{\Sigma}^{\oplus} \sigma d\tilde{\nu}(\sigma) ~~.\]
 Assume that for 
 all measurable vector fields $\eta = (\eta_{\sigma})_{\sigma \in \Sigma}
 \in {\cal H}_{\tilde{\pi}}$
 fulfilling $\eta_{\sigma} \in {\rm dom~} (K_{\sigma}^{-1/2})$
 $\tilde{\nu}$-a.e., 
 the admissibility criterion
 \[ \eta ~\mbox{admissible for }\tilde{\pi} \Longleftrightarrow
 \| K_{\sigma}^{-1/2} \eta_\sigma \|
 = 1 ~\tilde{\nu}\mbox{-almost everywhere} \]
 is valid. Then $\tilde{\nu} = \nu_G$ on $\Sigma$.
\end{prop}
\begin{prf}
 Denote by $T: \int_{\Sigma}^{\oplus} \sigma d\nu_G(\sigma) \to 
  \int_{\Sigma}^{\oplus} \sigma d\tilde{\nu}(\sigma) $
 the unitary intertwining operator obtained by
 \[ T((\phi_{\sigma})_{\sigma \in \Sigma}) = \left(\sqrt{\frac{d \tilde{\nu}}{
 d \nu_G}(\sigma)} \phi_{\sigma} \right)_{\sigma \in \Sigma} ~~.\]
 Clearly, a unitary equivalence maps admissible vectors onto 
 admissible vectors. Moreover,
 if $\eta = (\eta_{\sigma})_{\sigma \in \Sigma}$ fulfills
 $\eta_{\sigma} \in {\rm dom~} (K_{\sigma}^{-1/2})$ $\nu_G$-almost
 everywhere, then $T \eta$ does as well. Hence, if we let
 $\eta \in {\cal H}_{\pi}$ be any vector fulfilling the
 admissibility criterion from Theorem \ref{adm_crit_mfree},
 the necessary admissibility criterion for $\tilde{\pi}$ gives
 \[ \| K_{\sigma}^{-1/2} \eta_{\sigma} \| = 1 =
 \left\|\sqrt{\frac{d \tilde{\nu}}{
 d \nu_G}(\sigma)} K_{\sigma}^{-1/2} \eta_{\sigma} \right\| 
 \mbox{ almost everywhere }~~.\] 
 Hence the Radon-Nikodym-derivative is 1 almost everywhere.
\end{prf}

\section{Concrete and abstract admissibility conditions}

Now we are in a position to establish the connection between the
results in Sections 1 and 2.
In order to apply the theorems from section 2, let us assume that the 
semidirect product has a type-I regular representation.
In addition, we restrict attention to the case that
$\rhkdu/H$ is standard.

The first and crucial step consists in writing $\pi$ as the direct 
integral of monomial representations, as described in the following 
result (which is \cite[Theorem 2.1]{Li}). The direct integral
decomposition is obtained by looking at the dual representation
(\ref{dual_rep}) in the light of the measure decomposition from
Lemma \ref{lem_meas_disint} (a). Note that if $\rhkdu/H$ is
standard, the measure disintegration described in \ref{lem_meas_disint}
(a) can be given for all of $\rhkdu$ instead of $\Omega_{rc}$,
i.e., $\overline{\lambda}$ is given on $\rhkdu/H$, and
the $\beta_{\gamma H}$ exist also for orbits $\gamma H$
without compact stabilizer.

\begin{lemma}
 The quasiregular representation $\pi$ is the direct integral
 \[ \pi \simeq \int_{\rhkdu{}/H}^{\oplus} ({\rm Ind}_{G_{\gamma}}^G \gamma
 \times 1) d\overline{\lambda} (\gamma H), \]
 of irreducible representations, where $G_{\gamma} = \rhk \semdir H_{\gamma}$
 and $H_{\gamma}$ is the ``little fixed group'' of the character $\gamma$.
 In particular $\pi$ is multiplicity-free.
\end{lemma}
\begin{prf}
 The disintegration
 $d\lambda(\omega) = d\beta_{\gamma H} (\omega) d\overline{\lambda}
 (\gamma H)$ allows to interpret the conjugated representation $\widehat{\pi}$
 as a direct integral over $\rhkdu /H$ of representations acting on the spaces
 $\lpraum{2}{\gamma H, \mu_{\gamma H}}$,
 by identifying each $\widehat{f}$ with the family of restrictions
 $(\widehat{f}|_{\gamma H})_{\gamma H \in \rhkdu / H}$.
 The representations over the orbits are easily seen to be induced
 in the indicated manner, using a concrete
 realization of ${\rm Ind}_{G_{\gamma}}^G \gamma \times 1$ on 
 $\lpraum{2}{\gamma H, \mu_{\gamma H}}$ via cross sections.
 The remaining statements are immediate consequences of Mackey's theory.
\end{prf}

An algorithm for the construction of Plancherel measures for 
semidirect product was
provided by Kleppner and Lipsman in the fundamental papers
\cite{KlLi} I, II. Let us give an outline:
By Mackey's theory, the dual of $G$ may be described by 
\[ \widehat{G} = \bigcup_{\gamma H \in \rhkdu /H} \{ {\rm Ind}_{G_{\gamma}}^G 
 \gamma \times \sigma : \sigma \in \widehat{H}_{\gamma} \},
\]
whenever $\rhkdu /H$ is standard.
An intuitive interpretation is that $\widehat{G}$ may be considered as a 
``fibred space'', with 
base space $\rhkdu /H$ and the fibre over $\gamma H$ is (identified with)
$\widehat{H}_{\gamma}$,
and it is the central result of \cite{KlLi} (see part II, Theorem 2.3) 
that the Plancherel measure is obtained as a ``fibred measure'' as well:
Take any pseudo-image $\overline{\lambda}_{\nu}$ of standard Lebesgue 
measure, choose
Plancherel measures $\nu_{\gamma H}$ of the little groups $H_{\gamma}$
which exist if we assume that almost every $H_{\gamma}$ has a type-I
regular representation. 
Then the measure $\nu$ defined by
\begin{equation} \label{fib_meas}
 \nu (A) = \int_{\rhkdu /H} \int_{\widehat{H}_{\gamma}}
 \chi_A ( {\rm Ind}_{G_{\gamma}}^G \gamma \times \sigma)
 d\nu_{\gamma H}  (\sigma) d\overline{\lambda}_{\nu} (\gamma H) \end{equation}
is equivalent to the Plancherel measure. This determines
the measure class. If $G$ is unimodular, the proof of
\cite[II, Theorem 2.3]{KlLi} in fact gives a recipe how to obtain
the correct normalization; whereas in the non-unimodular case
the results of \cite{KlLi} only give the measure class, and no
access to the Duflo-Moore operators.

The ``fibred measure''-view of Plancherel measure provides a neat
interpretation of the set $\Omega_{c}$. As calculated in the previous
lemma, the 
quasi-regular representation is a direct integral, and the measure
defining it is supported on a subset of $\widehat{G}$ which meets each 
fibre in precisely one point, that is in $\rm Ind_{G_{\gamma}}^G \gamma \times
1$, which corresponds to the trivialrepresentation in $\widehat{H_{\gamma}}$.
The Plancherel measure of the trivial representation
in $\widehat{H_{\gamma}}$ is positive if and only if
the $H_{\gamma}$ is compact. Hence the inner integral in
(\ref{fib_meas}) vanishes at these points, and the part of the support
corresponding to the complement of $\Omega_c$ is a Plancherel-null set.
On the other hand, the construction of Plancherel measure shows that 
on the part of the support corresponding to $\Omega_c$, Plancherel
measure and $\overline{\lambda}$ are equivalent. Indeed, we are
free to take $\overline{\lambda_{\nu}}= \overline{\lambda}$, and the 
inner integral in (\ref{fib_meas}) is positive if $H_{\gamma}$ is compact.
In short, $\rhkdu = \Omega_c \cup (\rhkdu \setminus \Omega_c)$
corresponds to the decomposition of $\overline{\lambda}$
into a part which is absolutely continuous with respect to Plancherel
measure, and a part which is singular. This yields an abstract
explanation of the role of $\Omega_{c}$, in particular of
the necessary condition in Theorem \ref{adm_crit_elem}.


Before we describe the relationship between concrete and
abstract admissibility conditions, it is useful to relate
the Borel space $\Omega_{rc}/H$ to a suitable subset of $\widehat{G}$.

\begin{prop}
\label{Borel_iso}
 There exists a conull, $H$-invariant subset
 $U_0 \subset \Omega_{rc}$, such that the mapping
 \[ \Phi: U_0/H \ni \gamma H \mapsto {\rm Ind}_{G_\gamma}^G \gamma \times 1 \]
 is a Borel isomorphism onto a standard measurable subset
 $\Sigma \subset \widehat{G}$.
\end{prop}

\begin{prf}
 We just noted that $\pi_U$ is multiplicity-free with
 \[ \pi_U  \simeq \int_{U/H}^{\oplus} ({\rm Ind}_{G_{\gamma}}^G \gamma
 \times 1) d\overline{\lambda} (\gamma H). \]
 On the other hand,
 the existence of a weakly admissible vector for $\pi_U$ implies that
 $\pi_U$ is equivalent to a subrepresentation of the regular representation.
 This gives the alternative decomposition
 \[ \pi_U \simeq \int_{\Sigma_1}^{\oplus} \sigma d\nu_G(\sigma) ~~.\]
 Hence we may invoke \cite[Theorem, p.117]{Ma} to see that $T$ arises
 from an isomorphism
 of the underlying Borel spaces: There exist conull subset $\Sigma \subset
 \Sigma_1$, $U_0 \subset \Omega_{rc}/H$, a Borel isomorphism
 $\Phi : U_0 \to \Sigma$, a measurable field of unitary operators
 $T_{\gamma H} : {\rm L}^2(\gamma H,d\beta_{\gamma H}) \to {\cal H}_{
 \Phi(\gamma H)}$
 and a Radon-Nikodym derivative $\Psi: U_0/H \to \RR_0^+$ such
 that $T$ decomposes into $(\Psi(\gamma H) T_{\gamma H})_{\gamma H}$.
 Since $T$ is an intertwining operator, so is almost every
 $T_{\gamma H}$, and thus $\Phi(\gamma H) = {\rm Ind}_{G_{\gamma}}^G
 (\gamma \times 1)$.
\end{prf}

Let us now summarize the transfer between concrete and abstract
admissibility conditions.

\begin{thm}
 Let $\Phi$, $U_0$ and $\Sigma_0$ be as in Proposition \ref{Borel_iso}.
 For $\gamma H \subset U_0$ let ${\cal K}_{\gamma H}$ denote the
 operator on ${\rm L}^2(\gamma H,d\beta_{\gamma H})$
 given by pointwise multiplication with $\kappa|_{\gamma H}$.
 $\Phi$ gives rise to the following correspondences between the
 objects in Section 1 and those in Section 2:
 \begin{eqnarray*}
 U_0 / H & \longleftrightarrow & \Sigma_0 ~~,\\
  \gamma H & \longleftrightarrow & \sigma  ~~,\\
 {\rm L}^2(\gamma H, d\beta_{\gamma H})  & \longleftrightarrow &
 {\cal H}_{\sigma}  ~~, \\
  \widehat{f}|_{\gamma H} & \longleftrightarrow & \eta_{\sigma}  ~~,\\
 S_H(\widehat{f}) (\gamma H)  & \longleftrightarrow & \| \eta_{\sigma} \|
 ~~,\\
 \overline{\lambda} & \longleftrightarrow & \nu_G ~~,\\
  {\cal K}_{\gamma H} & \longleftrightarrow & K_{\sigma}  ~~,\\
 T_H(\widehat{f}) (\gamma H) & \longleftrightarrow &
 \| K_{\sigma}^{-1/2} \eta_{\sigma} \| ~~.
 \end{eqnarray*}   
 These correspondences exhibit Theorems \ref{adm_crit_elem} and 
 \ref{ex_ad_vec_qr} as special instances of Theorem \ref{adm_crit_mfree}.
\end{thm}

\begin{prf}
 It remains to check
 that the Duflo-Moore $K_{\sigma}$ corresponds to ${\cal K}_{\gamma H}$,
 and that the Plancherel measure $\nu_G$ belonging to this
 particular choice of Duflo-Moore operators corresponds to
 $\overline{\lambda}$.
 Straightforward calculation, using relation
 (\ref{sem_inv_konk}) from Lemma \ref{lem_meas_disint}, shows that
 ${\cal K}_{\gamma H}$ satisfies the quasi-invariance relation
 \[
 \left( {\rm Ind}_{G_{\gamma}}^G(\gamma \times 1) (x,h) \right)~
 {\cal K}_{\gamma H}
  ~ \left( {\rm Ind}_{G_{\gamma}}^G(\gamma \times 1) (x,h) \right)^* = 
 \Delta_G(x,h)^{-1} {\cal K}_{\gamma H}~~.  
 \]
 By \cite[Corollary 1 to Theorem 5]{DuMo}, the same relation
 has to be fulfilled by the Duflo-Moore operators; in fact
 by \cite[Lemma 1]{DuMo}, the relation characterizes the Duflo-Moore
 operators up to a scalar multiple. Since in addition, the measurability
 of $\kappa$ ensures that $({\cal K}_{\gamma H})_{\gamma H \in U_0/H}$
 is measurable, we may take this operator field as a realization
 of the Duflo-Moore operators.

 Then it remains to check is that, given this particular choice of
 Duflo-Moore operators, the measure $\overline{\lambda}$ is the
 corresponding Plancherel measure. But this is provided by
 the concrete admissibility condition
 in Theorem \ref{adm_crit_elem} together with Proposition \ref{adm_and_pm}.
\end{prf}

\section*{Concluding remarks}
 
 The use of discrete dilation groups can be seen as a first 
 discretization step; for a fully discrete wavelet transform
 we would have to discretize the translations as well.
 Recent publications \cite{Ba,Woj,LPT} document the increasing interest in the
 use of direct integral decompositions for the study and construction
 of fully discrete wavelet (or wavelet-like) systems. We expect
 that a connection between our results and the results contained
 in these papers would
 provide a better understanding of the discretization problem.

 Our main motivation was to study admissibility conditions
 both by more or less elementary methods and in connection
 with Plancherel theory. Our paper can be seen as a natural
 continuation of the paper by Bernier and Taylor. Our construction
 of Duflo-Moore operators is entirely analogous to what is done in
 \cite{BeTa} for the case of open free orbits. Irreducible square-integrable
 subrepresentations of $\pi$ correspond to open orbits, which appear
 as atoms in the quotient space $\Omega_{rc}/H$.
 The increase in complexity that comes from considering reducible 
 representations results in the transition from
 counting measure on a finite set of open orbits to a certain measure
 on the quotient space. In addition, the issue of proper normalization
 comes up. 

 As we mentioned in the Introduction, Plancherel theory did not
 help establish admissibility conditions or construct
 admissible vectors for the quasi-regular representation.
 It was rather the other way round:
 The close connection between the admissibility conditions and
 Plancherel measure, as observed in Proposition \ref{adm_and_pm},
 turned out to be a useful tool for the explicit calculation of Plancherel
 measure and the Duflo Moore operators. Nevertheless, given
 an arbitrary reducible representation, the connection to
 Plancherel theory can provide an orientation and a possible
 strategy for the construction of admissible vectors.
 We believe that the semidirect products
 studied in this paper are very well suited to explicitly see
 these aspects of Plancherel theory at work.

 The results given here and in \cite{Fu3} could probably be
 generalized to hold for the type-I part of $\lambda_G$.
 Duflo and Moore devised their Plancherel theory for this
 more general setting, and we believe that the proofs in
 \cite{Fu3} should go through as well. 
 Similarly, the requirement that $\rhkdu / H$ be standard
 could probably be replaced by something weaker. Generally
 speaking, the type-I requirement is not needed for the
 existence of admissible vectors. For instance, admissible
 vectors exist for the regular representation of an arbitrary discrete
 group (simply take the $\delta$ at the origin), but only 
 very few of these groups have a type-I regular representation;
 confer \cite{K}. Also, as examples in the literature illustrate, 
 Plancherel measure can exist in the non-type-I setting,
 but it is no longer unique. This will probably have consequences
 for the necessary condition in Theorem \ref{ex_adm_vec_abs}(a)
 for the existence of admissible vectors. However, for the
 construction of admissible vectors for the regular representation
 of a non-unimodular group, it is conceivable that any Plancherel
 decomposition might
 work. (Note however that the construction in \cite{Fu3} rests
 on the concrete description of the Duflo-Moore operators in \cite{DuMo},
 which is valid only for the type-I setting.)

 Despite the discussion for the general case, the existence
 of admissible vectors for the quasiregular representation
 (or some subrepresentation) seems to be tied more closely to
 a regularity condition on the orbit space, which in turn
 is be related to the type of the regular representation.
 In this context there is essentially one open question left, and that regards
 the role of the set $\Omega_c$, or rather, $\Omega_c \setminus 
 \Omega_{rc}$. As we have seen in Example \ref{c_ex_Omega_c},
 the existence of admissible vectors corresponding to 
 subsets of the latter set is not guaranteed.
 The following conjecture, which is a sharpening of a statement from
 Theorem \ref{adm_crit_elem}, would neatly resolve the question;
 unfortunately discrete dilation groups acting ergodically
 (such as in Example \ref{c_ex_Omega_c}) are so far our only
 evidence for its truth. However, in the observations following
 \cite[Proposition 2.8]{LWWW} the authors make a conjecture
 similar to the one we give here.

\vspace{0.5cm} 
\noindent
 {\bf Conjecture:} If $\pi_U$ has an admissible vector, then
 $U \subset \Omega_{rc}$.

\begin{appendix}
 
\section{The sets $\Omega_c$ and $\Omega_{rc}$}

In this appendix we prove the measurability of $\Omega_c$ and
the openness of $\Omega_{rc}$. The proof for the first result uses the 
subgroup space of $H$, as introduced by Fell \cite{Fe}.

\begin{defrem}
 Let $G$ be a locally compact group. The {\bf subgroup space of $\mathbf{G}$}
 is the set $K (G) := \{ L < G: L \mbox{ is closed } \} $, endowed with
 the topology generated by the sets 
 \[ U (V_1,\ldots,V_n;C) := \{ L \in K(G) : L \cap V_i \not= \emptyset,
 \forall 1 \le i \le n, L \cap C = \emptyset \}, \]
 where $V_1,\ldots,V_n$ denotes any finite family of open subsets of $G$
 and $C \subset G$ is compact.\\
 With this topology $K(G)$ is a compact Hausdorff space.
\end{defrem}

The next few results are probably well known, but we were not able to
find references for them:
\begin{prop}
\label{K(G)_sec_ct}
 If $G$ is second countable, then so is $K(G)$.
\end{prop}
\begin{prf}
 Clearly it suffices to consider only $U(V_1,\ldots,V_n;C)$ with 
 $V_i$ belonging to a countable base of the topology on $G$, hence
 we are done when we show that for each $C \subset G$ compact
 we have $U(G;C) = \bigcup_{n \in \NN} U(G;C_n)$, with the $C_n$ belonging
 to a fixed countable collection of compact sets. For this purpose,
 let $A \subset G$ be countable and dense and let ${\cal U}$
 be a countable neighborhood base of unity consisting of compact sets.
 Then, for a given compact $C$ and a closed subgroup $H$ with $H \in U(G;C)$
 there exist $a_1,\ldots,a_m \in A$ and $U_1,\ldots,U_m
 \in {\cal U}$ with $C \subset \bigcup_{i=1}^m a_i U_i \subset G \setminus H$.
 Hence $H \in U(G;\bigcup_{i=1}^m a_i U_i) \subset U(G;C)$, which shows the
 claim.  
\end{prf}

\begin{prop}
\label{int_meas}
 Let $G$ be a Lie group and $H < G$ be closed. Then the intersection mapping
 \[
 {\cal I}: K (G) \to K(H), L \mapsto L \cap H 
 \]
 is Borel.
\end{prop}
\begin{prf}
 Since $H$ is second countable, the sets of the type $U(V;\emptyset)$ ($V \subset H$
 open) and $U(H;C)$ ($C \subset H$ compact) generate the Borel structure of
 $K(H)$, hence we need only take care of these. Clearly ${\cal I}^{-1} (U(H;C)) =
 \{ L \in K(G) :  L \cap C = \emptyset \}$ is open. Now let $V \subset H$
 be open, $V = \tilde{V} \cap H$ for some open set $\tilde{V} \subset G$.
 Since $G$ is a Lie group, $\tilde{V} = \bigcup_{i \in \NN} F_i$ with
 closed subset $F_i$. Furthermore $H = \bigcup_{j \in \NN} C_j$ with compact
 sets $C_j$. Hence, 
 \begin{eqnarray*}
 {\cal I}^{-1} (U(V;\emptyset)) & = & \{ L \in K(G) : L \cap H \cap \tilde{V} \not= 
 \emptyset \} \\
 & = & \bigcup_{i,j \in \NN} \{ L \in K(G) : L \cap C_j \cap F_i \not=
 \emptyset \}
 \end{eqnarray*}
 is a countable union of closed sets, since the sets $C_j \cap F_i$ are
 compact. 
\end{prf}

Now we can consider the stabilizer mapping $\gamma \mapsto H_{\gamma}$:
\begin{prop}
 Let $H < {\rm GL}(k,\RR)$ be closed. Then the stabilizer
 mapping $\rhkdu \ni \gamma \mapsto H_{\gamma} \in K(H)$ is 
 a Borel mapping. The set $\Omega_c$ is a Borel subset of $\rhkdu$.
\end{prop}
\begin{prf}
 By \ref{int_meas} it is sufficient to consider $H = {\rm GL}(k,\RR)$.
 For this case fix $\gamma_1 \in \rhkdu{} \setminus \{ 0 \}$ and
 let $\rho : \rhkdu{}  \setminus \{ 0 \} \to H$ be any
 measurable cross section, i.e., $\rho$ fulfills $\gamma_1 \rho (\omega) =
 \omega$ for all $\omega \in \rhkdu \setminus \{ 0 \}$. Then we have
 $H_{\omega} = \rho (\omega) H_{\gamma_1} \rho (\omega)^{-1}$,
 hence the stabilizer mapping equals $c \circ \rho$, with
 $c (x) := x H_{\gamma_1} x^{-1}$. The mapping $c:H \to K(H)$ is
 easily seen to be continuous, hence the stabilizer mapping is measurable.

 For the last statement it suffices to show that $K_c(H) := \{ L \in K(H):
 L \mbox{ is compact} \}$ is a Borel subset of $K(H)$. For this purpose
 let $(C_n)_{n \in \NN}$ be a countable family of compact subsets of $H$ with
 the property that for each $K \subset H$ compact there exists
 an $n \in \NN$ with $K \subset C_n$. Such a family was constructed in
 the proof of Proposition \ref{K(G)_sec_ct}. We then have
 \[
  K_c(H)  = \bigcup_{n \in \NN} \{ L \in K(H) : L \cap (H \setminus C_n) =
 \emptyset \}, 
 \]
 whence we see that $K_c(H)$ is the countable union of closed sets.
\end{prf}

The proof of the following proposition uses ideas from
\cite{LWWW}.
\begin{prop} $\Omega_{rc}$ is open.
\end{prop}
\begin{prf}
Define 
\[ H_{\omega}^{\epsilon} = \{ h \in H : | \omega h - \omega | \le \epsilon
 \}~~, \]
where $| \cdot |$ denotes the euclidean norm on $\rhkdu$.
If $H_{\omega}^{\epsilon}$ is compact for some $\epsilon>0$, then
$B_{\epsilon}(\omega) \cap \omega H = \omega H_{\omega}^{\epsilon}$
is compact. (Here $B_{\epsilon}(x)$ denotes the closed $\epsilon$-ball
around $x$.) Hence the orbit $\omega H$ is locally closed.
Conversely, assume that $B_{\epsilon}(\omega) \cap \omega H$ is 
compact for some $\epsilon>0$ and that $H_{\omega}$ is compact. 
There exists a measurable cross-section $\tau: \omega H \to H$ which maps 
compact sets in $\omega H$ to relatively compact sets in $H$. 
Hence $H_{\omega}^{\epsilon} \subset H_\omega \tau(B_{\epsilon}(\omega))$
is relatively compact and closed, hence compact.
In short, we have shown
\[ \omega \in \Omega_{rc} \Longleftrightarrow \exists \epsilon>0 :
 H_{\omega}^{\epsilon} ~\mbox{is compact }~~, \]
and we are going to use this characterization to prove the
openness of $\Omega_{rc}$. 

If the origin is in $\Omega_{rc}$,
then $H$ is compact, and $\Omega_{rc} = \rhkdu$.
In the other case, pick $\omega$ in the complement and 
$\epsilon>0$ with $H_{\omega}^{\epsilon}$ compact. Since
${\rm GL}(k,\RR)$ acts transitively on $R^n \setminus 0$, we may
(possibly after passing to a smaller $\epsilon$) 
assume that there exists a continuous cross-section
$\sigma : B_{\epsilon} (\omega) \to {\rm GL}(n,R)$ with relatively
compact image, i.e., $\omega \sigma(\gamma) = \gamma$, for all 
$\gamma \in B_{\epsilon}(\omega)$, and $\sigma(B_{\epsilon}(\gamma))
\subset U$,
where $U$ is a compact neighborhood of the identity in
${\rm GL}(k,\RR)$. We are going to show that $B_{\epsilon}(\omega) \subset \Omega_0$.
For this purpose let $\gamma \in B_{\epsilon}(\omega)$. Clearly
it is enough to prove that
\[ C := \{ h \in H : \gamma h \in B_{\epsilon}(\omega) \} = 
 \{ h \in {\rm GL}(k,\RR) : \gamma h \in B_{\epsilon}(\omega) \} \cap D \] 
is relatively compact. By assumption,
\[ H_{\omega}^{\epsilon} = \{ h \in {\rm GL}(k,\RR) : \omega H \in B_{\epsilon}(
 \omega) \} 
 \cap D \]
is compact. Hence
\begin{eqnarray*}
 C & = & \{ h \in {\rm GL}(k,\RR) : \omega \sigma(\gamma) h \in B_{\epsilon}(
 \omega) \} \cap
 D \\
 & = & \sigma(\gamma)^{-1}  \{ h \in {\rm GL}(k,\RR) : \omega h \in B_{\epsilon}
 (\omega) \} \cap D \\
 & \subset &  U^{-1} ( \{ h \in {\rm GL}(k,\RR) : \omega h \in B_{\epsilon}(\omega)
 \} \cap D)
\end{eqnarray*}
i.e., $C$ is contained in the product of two compact sets, and thus
relatively compact. (Note that we used here that $H$ is a closed
subgroup of ${\rm GL}(k,\RR)$, hence compactness in $H$ is the same as
compactness in ${\rm GL}(k,\RR)$.)
\end{prf}

\end{appendix}

\section*{Acknowledgements} We would like to thank Guido Weiss
 and his collaborators for making the manuscripts \cite{WW}
 and \cite{LWWW} available to us. Thanks are also due to S. Twareque Ali
 and Bruno Torr\'esani for suggestions for the final version of
 this paper.

\end{document}